\documentclass[aps,prl,twocolumn,showpacs,superscriptaddress,groupedaddress]{revtex4}  % for review and submission
\usepackage{graphicx}  % needed for figures
\usepackage{dcolumn}   % needed for some tables
\usepackage{bm}        % for math
\usepackage{amssymb}   % for math

\usepackage{epsf}

\usepackage{graphicx}

\newcommand{\insertplot}[5]{\begin{figure}
 \hfill\hbox to 0.05in{\vbox to #5in{\vfill
 \inputplot{#1}{#4}{#5}}\hfill}
 \hfill\vspace{-.1in}
 \caption{#2}\label{#3}
 \end{figure}}
 \newcommand{\inputplot}[3]{
 \special{ps: plotfile #1}
}
\newcounter{fig}   

\newcommand{\vphi}{\varphi}

\usepackage{graphicx}

\begin{document} 
\title{
Sequences of extremal radially excited rotating black holes}
 \vspace{1.5truecm}
\author{
{\bf Jose Luis Bl\'azquez-Salcedo$^1$},
{\bf Jutta Kunz$^2$},\\
{\bf Francisco Navarro-L\'erida$^3$},
{\bf Eugen Radu$^2$}\\
%
%\vspace{1.5truecm}
%
$^1$
Dept.~de F\'{\i}sica Te\'orica II, Ciencias F\'{\i}sicas\\
Universidad Complutense de Madrid, E-28040 Madrid, Spain\\
$^2$ Institut f\"ur  Physik, Universit\"at Oldenburg\\ Postfach 2503,
D-26111 Oldenburg, Germany\\
$^3$
Dept.~de F\'{\i}sica At\'omica, Molecular y Nuclear, Ciencias F\'{\i}sicas\\
Universidad Complutense de Madrid, E-28040 Madrid, Spain
}

%\vspace{1.5truecm}
\pacs{04.40.Nr, 04.50.-h, 04.20.Jb} 
\date{\today}

%\vspace{1.0truecm}

\begin{abstract}
In Einstein-Maxwell-Chern-Simons theory 
the extremal Reissner-Nordstr\"om solution is no longer the single
extremal solution with vanishing angular momentum,
when the Chern-Simons coupling constant reaches a critical value.
Instead a whole sequence of rotating extremal $J=0$ solutions arises,
labeled by the node number of the magnetic U(1) potential. 
%enumerated by the node number of their functions. 
Associated with the same near horizon solution,
the mass of these radially excited extremal solutions 
converges to the mass of the extremal Reissner-Nordstr\"om solution.
%which, however, is associated with a different near horizon solution.
%In the critical case, we deduce an infinite non-uniqueness 
%for the extremal black holes.
On the other hand, not all near horizon solutions 
are also realized as global solutions.

%while other near horizon solutions correspond 
%to a sequence of global solutions.
\end{abstract}

\maketitle

%\vfill\eject
  \medskip
%%%%%%%%%%%%%%%%%%%%%%%%%%%%%%%%%%%%%%%%%%%%%%%%%%%%%%%%%%%%%%%%%%%
%\section{Introduction}
%%%%%%%%%%%%%%%%%%%%%%%%%%%%%%%%%%%%%%%%%%%%%%%%%%%%%%%%%%%%%%%%%%%
%%%%%%%%%%%%%%%%%%%%%%%%%%%%%%%%%%%%%%%%%%%%%%%%%%%%%%%%%%%%%%%%%%%%%%%%%%%%%%%
{\bf Introduction.--}
%%%%%%%%%%%%%%%%%%%%%%%%%%%%%%%%%%%%%%%%%%%%%%%%%%%%%%%%%%%%%%%%%%%%%%%%%%%%%%%%
Higher-dimensional black hole spacetimes
have received much interest in recent years,
associated with
various developments in gravity and high energy physics.
In particular, the first successful
statistical counting of black hole entropy in string theory
was performed for an extremal static Reissner-Nordstr\"om (RN)
black hole in five spacetime dimensions  \cite{Strominger:1996sh}.

However, in odd dimensions the Einstein-Maxwell (EM) action may be supplemented by a
Chern-Simons (CS) term. In 5 dimensions, for a certain value of the CS 
coefficient $\lambda=\lambda_{\rm SG}$, 
the theory corresponds to the bosonic sector of $D=5$ supergravity,
where rotating black hole solutions are known analytically 
\cite{Breckenridge:1996sn,Breckenridge:1996is,Cvetic:2004hs,Chong:2005hr}.
A particular interesting subset of these black holes,
the BMPV \cite{Breckenridge:1996is} solutions, corresponds
to extremal cohomogeneity-1 solutions, where both angular momenta
have equal magnitude, $|J_1|=|J_2|=|J|$.
%Their horizon angular velocities $ \Omega_{\rm H}$ vanish.
These black holes have a non-rotating horizon, although their angular momentum
 is nonzero.
%In these ergo-region free black holes
It is stored in the Maxwell field,
with a negative fraction of the total angular momentum
stored behind the horizon 
\cite{Gauntlett:1998fz,Herdeiro:2000ap,Townsend:2002yf}.
%With increasing angular momentum, the horizon becomes
%increasingly squashed, reaching a singular solution
%with vanshing area for a maximal value of $J$.

As conjectured in \cite{Gauntlett:1998fz},
supersymmetry is associated with the borderline 
between stability and instability,
since for $\lambda>\lambda_{\rm SG}$ a rotational instability arises,
where counterrotating black holes appear \cite{Kunz:2005ei}.
Moreover, when the CS coefficient is increased beyond the 
critical value of $2\lambda_{\rm SG}$, EMCS
black holes - with the horizon topology of a sphere -
are no longer uniquely characterized by their global charges 
\cite{Kunz:2005ei}.

Focussing
on extremal solutions with equal
magnitude angular momenta,
we here reanalyze 5-dimensional EMCS black holes  in the vicinity
and beyond the critical value of the CS coupling constant $\lambda_{\rm SG}$.
We obtain these cohomogeneity-1 solutions numerically,
solving the field equations with appropriate boundary conditions.

These extremal black holes are associated with analytical near horizon solutions,
obtained in the entropy function formalism.
Surprisingly, however, certain sets of near horizon solutions
are associated with more than one global solution,
whereas other sets of near horizon solutions
do not possess global counterparts.
In particular, we find whole sequences of radially excited
extremal solutions, all with the same area and angular momenta 
for a given charge.

%%%%%%%%%%%%%%%%%%%%%%%%%%%%%%%%%%%%%%%%%%%%%%%%%%%%%%%%%%%%%%%%%%%
%\section{Action, Ansatz, Charges}
%%%%%%%%%%%%%%%%%%%%%%%%%%%%%%%%%%%%%%%%%%%%%%%%%%%%%%%%%%%%%%%%%%%
%%%%%%%%%%%%%%%%%%%%%%%%%%%%%%%%%%%%%%%%%%%%%%%%%%%%%%%%%%%%%%%%%%%%%%%%%%%%%%%
{\bf The model.--}
%%%%%%%%%%%%%%%%%%%%%%%%%%%%%%%%%%%%%%%%%%%%%%%%%%%%%%%%%%%%%%%%%%%%%%%%%%%%%%%%
We consider the EMCS
action with Lagrangian \cite{Gauntlett:1998fz}
\begin{eqnarray}
{\cal L}= \frac{1}{16\pi G_5} [\sqrt{-g}(R - F^{2}
%F_{\mu \nu} F^{\mu \nu}
)
- \frac{2\lambda}{3\sqrt{3}}\varepsilon^{\mu\nu\alpha\beta\gamma}A_{\mu}F_{\nu\alpha}F_{\beta\gamma}] ,
\label{Lag}
\end{eqnarray}
with curvature scalar $R$,
Newton's constant $G_5$,
gauge potential $A_\mu $,
field strength tensor
$ F_{\mu \nu} = \partial_\mu A_\nu -\partial_\nu A_\mu $,
and CS coupling constant ${ \lambda}$ (with $\lambda_{\rm SG}=1$).

To obtain stationary cohomogeneity-1 solutions
we employ for the metric the para\-metrization \cite{Kunz:2006eh}
\begin{eqnarray}
\label{metric}
&&ds^2 = -f(r) dt^2 + \frac{m(r)}{f(r)}(dr^2 + r^2 d\theta^2) \nonumber \\
&& + \frac{l(r)}{f(r)}r^2 \sin^2\theta \left( d \varphi -\frac{\omega(r)}{r}dt
\right)^2 \nonumber \\
\nonumber
&&+\frac{l(r)}{f(r)}r^2 \cos^2\theta \left( d \psi -\frac{\omega(r)}{r}dt \right)^2 
\nonumber \\ 
&&+\frac{m(r)-l(r)}{f(r)}r^2 \sin^2\theta \cos^2\theta(d \varphi  -d \psi)^2,
\end{eqnarray}
and for the gauge potential
\begin{equation}
A_\mu dx^\mu  = a_0(r) dt + a_k(r) (\sin^2 \theta d\varphi+\cos^2 \theta d\psi).
\end{equation}
To obtain asymptotically flat solutions, the metric functions should satisfy
the following set of boundary conditions
at infinity:
$f|_{r=\infty}=m|_{r=\infty}=l|_{r=\infty}=1$,
$\omega|_{r=\infty}=0$.
For the gauge potential we choose a gauge such that
$a_0|_{r=\infty}=a_\vphi|_{r=\infty}=0$.
In isotropic coordinates the horizon is located at $r_{\rm H}=0$.
An expansion at the horizon yields
$f(r) = f_4 r^4 + f_{\alpha} r^{\alpha} + \dots$,
$m(r) = m_2 r^2 + m_{\beta} r^{\beta} + \dots$,
$l(r) = l_2 r^2 + l_{\gamma} r^{\gamma} + \dots$,
$\omega(r) = \Omega_H  r + \omega_2 r^2  + \dots$,
$a_0(r) = a_{0,0} + a_{0,\lambda} r^{\lambda}  + \dots$,
$a_k(r) = a_{k,0} + a_{k,\mu} r^{\mu}  + \dots$.
Interestingly, the coefficients $\alpha$, $\beta$, $\gamma$,
$\lambda$, $\mu$ and $\nu$ can be non-integer.

The global charges of these solutions can be read from the asymptotic expansion
\cite{Kunz:2005ei}
\begin{eqnarray}
 f=1-\frac{\ 8 G_5 M}{3 \pi r^{2}} + \dots \ ,~  
 \omega=\frac{4 G_5J}{\pi r^{3}}  + \dots \ ,   \nonumber \\ 
 a_0=-\frac{G_5Q}{\pi r^{2}} + \dots \ ,~   
 a_\vphi=\frac{G_5 { \mu}_{\rm mag}}{\pi r^{2}} + \dots \ ,
%
%f=1-\frac{G_5 M}{6 \pi^{2} r^{2}} + \dots \ , 
%\omega=\frac{G_5J}{4 \pi^{2} r^{3}}  + \dots \ ,  \nonumber \\ 
%a_0=-\frac{G_5Q}{4 \pi^{2} r^{2}} + \dots \ , 
%a_\vphi=- \frac{G_5 { \mu}_{\rm mag}}{4 \pi^{2} r^{2}} + \dots \ ,
\end{eqnarray}
together with their magnetic moment ${ \mu}_{\rm mag}$.
These extremal solutions satisfy the Smarr formula \cite{Gauntlett:1998fz}
%\begin{equation}
%\label{smarr}
%\frac{2}{3} M =  %\frac{\kappa A_{\rm H}}{8 \pi G_5} +
%2 \Omega_H J  +  \frac{2}{3} \Phi_{\rm H} Q  \ , 
$
  M = 3 \Omega_H J  +  \Phi_{\rm H} Q , 
$
%\end{equation}
and the first law 
%\begin{equation}
%\label{first-law}
$
dM =  2 \Omega_H dJ+\Phi_{\rm H} dQ    , 
$
%\end{equation}
where %$\kappa$ is the surface gravity, 
%$A_{\rm H}=2\pi^2 \sqrt{\frac{l_2}{f_4}}\frac{m_2}{f_4} $ 
$A_{\rm H}=2\pi^2 \sqrt{ {l_2} } {m_2}/{f_4}^{3/2} $ 
is the horizon area,
%(with $S=A_{\rm H}/4 G_5$ the entropy),
$\Omega_H$ is the horizon angular velocity, and $\Phi_{\rm H}=-(a_{0,0}+\Omega_H a_{k,0})$ is the horizon
electrostatic potential.
% (with $\Phi_{\rm H}=(a_0+\Omega_H a_k)|_{r_H}$).
%In extremal solutions $\kappa$ vanishes.

%%%%%%%%%%%%%%%%%%%%%%%%%%%%%%%%%%%%%%%%%%%%%%%%%%%%%%%%%%%%%%%%%%%
%\section{Near horizon formalism}
%%%%%%%%%%%%%%%%%%%%%%%%%%%%%%%%%%%%%%%%%%%%%%%%%%%%%%%%%%%%%%%%%%%
%%%%%%%%%%%%%%%%%%%%%%%%%%%%%%%%%%%%%%%%%%%%%%%%%%%%%%%%%%%%%%%%%%%%%%%%%%%%%%%
{\bf The near horizon solutions.--}
%%%%%%%%%%%%%%%%%%%%%%%%%%%%%%%%%%%%%%%%%%%%%%%%%%%%%%%%%%%%%%%%%%%%%%%%%%%%%%%%
A partial analytical understanding of the properties of the
solutions
can be achieved by studying their near horizon expression in conjunction
with the attractor mechanism \cite{Astefanesei:2006dd}. 
The advantage of the latter is that we can compute the physical charges and obtain semi-analytic
expressions for the entropy as a function of electric charge and angular momentum.
%empolying the attractor mechanism.
% in particular to express the
%entropy as a function of $J$ and $Q$. % to compute the entropy function
%compute their entropy functions
%The near horizon formalism needs special care, when it is employed in the
%presence of a CS term, as discussed extensively in the literature, see $e.g.$ \cite{Suryanarayana:2007rk}.

To apply the entropy function for the near horizon
geometry of the extremal EMCS solutions, one uses the ansatz \cite{Kunduri:2007qy}
\begin{eqnarray}
\label{metric_ansatz}
&&ds^2 = v_1(\frac{d\rho^2}{\rho^2}-\rho^2dt^2)  + v_2 
[
\sigma_1^2+\sigma_2^2+v_3 (\sigma_3 -\alpha \rho dt)^2
], \nonumber \\ && A_\mu dx^\mu =-e \rho dt +p \sigma_3,
%[ 4 d \theta^2+\sin^2 2\theta(d\phi_2-d\phi_1)^2]
%\\ 
%\nonumber
%&&+v_2\eta[d\phi_1+d\phi_2+\cos^2 2\theta(d\phi_2-d\phi_1)-\alpha \rho dt]^2,
\end{eqnarray}
(where $\sigma_1^2+\sigma_2^2=d\bar \theta^2+\sin^2\bar \theta d\psi^2$),
$\sigma_3=d\phi+\cos \bar \theta d \psi$, with $\bar \theta=2 \theta$, 
$\phi_1-\phi_2=\phi$, $\phi_1+\phi_2=\psi$),
and constants $v_a$,  $e$, $p$ and $\alpha$.
 
 %%%%%%%%%%%%%%%%%%%%%%%%%%%%%%%%%
\begin{figure}[t!]
\begin{center}
\mbox{\hspace*{-.8cm}
\includegraphics[height=.35\textheight, angle =270]{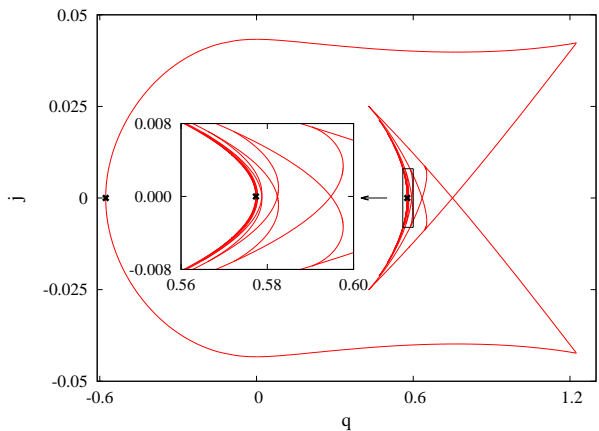}
}
\end{center}
\caption{Scaled angular momentum $j=J/M^{3/2}$  
versus the scaled electric charge $q=Q/M$
for extremal black holes ($\lambda=5$).
The asterisks mark the extremal static RN solutions.
%($Q=\pm 8 \sqrt{3} \pi^2$, $\lambda=5$).
}
\label{fig1}
\end{figure}
%%%%%%%%%%%%%%%%%%%%%%%%%%%%%%%%%

 %%%%%%%%%%%%%%%%%%%%%%%%%%%%%%%%%
\begin{figure}[t!]
\begin{center}
\mbox{\hspace*{-.8cm}
\includegraphics[height=.35\textheight, angle =270]{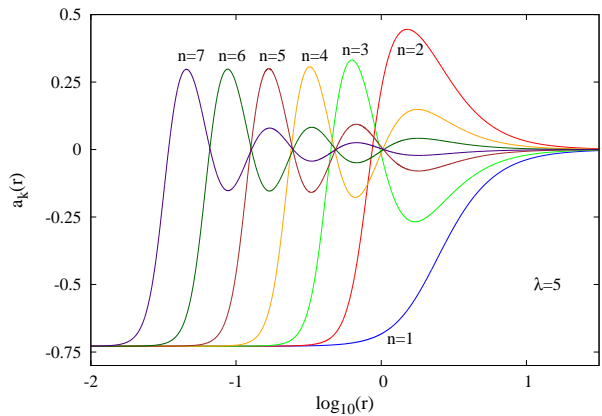} 
}
\end{center}
\caption{\small
The magnetic gauge potential $a_k(r)$ for rotating extremal solutions with $J=0$
and  the same electric charge 
Q,
%$Q=8 \sqrt{3} \pi^2$,
for several node numbers.
%($Q=\pm 8 \sqrt{3} \pi^2$, $\lambda=5$).
}
\label{fig2}
\end{figure}
%%%%%%%%%%%%%%%%%%%%%%%%%%%%%%%%%
 
  In the entropy function formalism, the entropy can be found from the extremum
of the entropy function $S=2\pi (\alpha J+e \hat q-f)$ in which 
$ f =\int d \theta d \varphi_1 d\varphi_2 \sqrt{-g} {\cal L}$
and $J=\partial f/\partial \alpha$, $\hat q=\partial f/\partial e$. 
However, the analysis is somehow intricate due
to the presence of the CS term.
% the constant $\hat q$ cannot be identified with 
%the electric charge.
For example,  for $\lambda \neq 0$,
the constant $\hat q$ cannot be identified with 
the electric charge 
and the extremization equations 
$\partial F/\partial v_a= \partial F/\partial e=\partial F/\partial \alpha=0$
should be used together with the Maxwell-Chern-Simons equations  \cite{Suryanarayana:2007rk}.
%, which contains an CS term contribution.
%By solving the equations 
%one finds 
The near horizon solution is found in terms of 
$p,v_1$ 
(this holds also for  $S,J$ and $Q$).
% (with $v_2=v_1$).
% and $v_3=2(p^2-e^2)/(\alpha^2-2)e^2+(2\alpha^2-1)p^2)$).
However, for a generic nonzero $\lambda \neq \lambda_{SC}$, 
it is not possible to  write an explicit expression of $S=A_H/4G_5$  as a function of $Q,J$.
Instead, a straightforward numerical study of the algebraic relations
reveals a rather complicated picture,
with several branches of solutions.
For example, two different near horizon solutions may exist with the same global charges $J,Q$ 
(see Fig.~\ref{fig4}).

%%%%%%%%%%%%%%%%%%%%%%%%%%%%%%%%%%%%%%%%%%%%%%%%%%%%%%%%%%%%%%%%%%%
%\section{Results}
%%%%%%%%%%%%%%%%%%%%%%%%%%%%%%%%%%%%%%%%%%%%%%%%%%%%%%%%%%%%%%%%%%%
%%%%%%%%%%%%%%%%%%%%%%%%%%%%%%%%%%%%%%%%%%%%%%%%%%%%%%%%%%%%%%%%%%%%%%%%%%%%%%%
{\bf The results.--}
%%%%%%%%%%%%%%%%%%%%%%%%%%%%%%%%%%%%%%%%%%%%%%%%%%%%%%%%%%%%%%%%%%%%%%%%%%%%%%%%
 The  global solutions are found by solving numerically the EMCS equations
 subject to the boundary conditions described above. 
In the numerical calculations \cite{numerics}, we introduce a compactified radial coordinate
$\bar{r}= r/(1+r)$   
and employ units such that $16 \pi G_5=1$.

\begin{figure}[t!]
\begin{center}
\mbox{\hspace*{-.8cm}
\includegraphics[height=.35\textheight, angle =270]{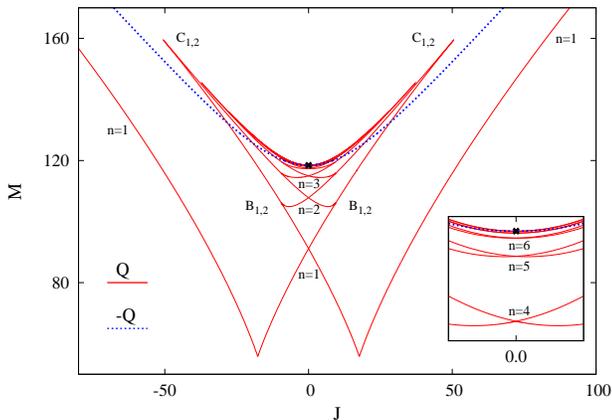}
}
\end{center}
\caption{Mass $M$ versus angular momentum $J$ 
for extremal black holes.
%  ($Q=\pm 8 \sqrt{3} \pi^2$, $\lambda=5$).
The asterisk marks the extremal static solution.
}
\label{fig3}
\end{figure}

\begin{figure}[t!]
\begin{center}
\mbox{\hspace*{-.8cm}
\includegraphics[height=.35\textheight, angle =270]{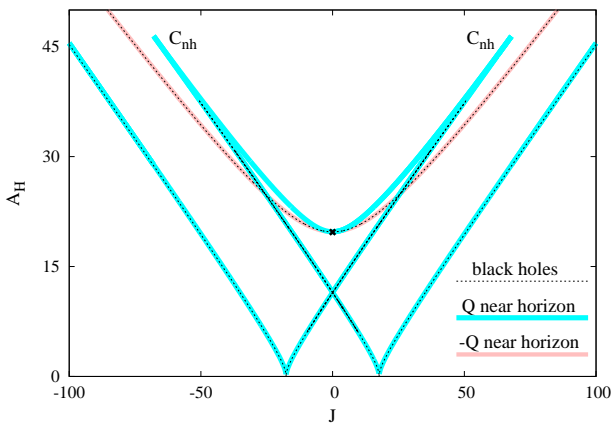}
}
\end{center}
\caption{\small
Same for the horizon area $A_{\rm H}$.
The near horizon solutions are also shown.
}
\label{fig4}
\end{figure}

\begin{figure}[t!]
\begin{center}
\mbox{\hspace*{-.8cm}
\includegraphics[height=.35\textheight, angle =270]{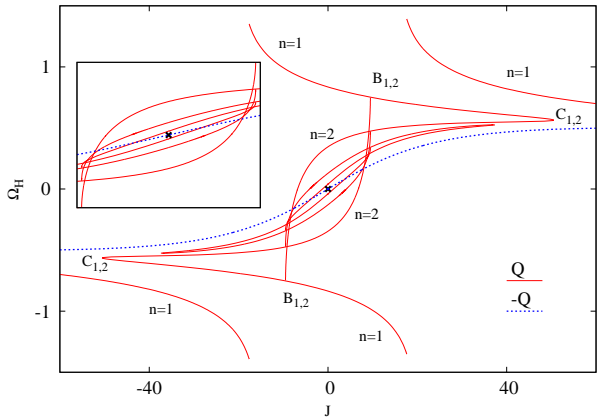}
}
\end{center}
\caption{Horizon angular velocity $\Omega_{\rm H}$ 
versus angular momentum $J$ 
for extremal black holes.
%  ($Q=\pm 8 \sqrt{3} \pi^2$, $\lambda=5$).
The asterisk marks the extremal static solution.
}
\label{fig5}
\end{figure}

We start  by exhibiting in Fig.~\ref{fig1} 
the scaled angular momentum $j=J/M^{3/2}$ versus
the scaled charge $q=Q/M$.
The results there and also in Figures \ref{fig3}, \ref{fig4} and \ref{fig5} are for a
CS coupling constant $\lambda=5$; however, a similar picture has been found for other values of $\lambda>2$.
% for extremal black hole solutions at CS coupling $\lambda=5$.
Fig.~\ref{fig1} exhibits the domain of existence of EMCS black holes,
since all non-extremal black hole solutions reside within the outer boundary, formed
by  extremal black holes.

For spinning solutions, the CS term breaks the symmetry $Q\to -Q$.
The extremal solutions with negative charge  form the left outer boundary.
This contains the extremal static solution where $J$ vanishes.
The extremal solutions with positive charge, on the other hand, 
represent a much more interesting set of solutions.

%%%%%%%%%%%%%%%%%%%%%%%%%%%%%%%%%%%%%%
%%%%%%%%%%%%%%%%%%%%%%%%%%%%%%%%%%%%%%

%%%%%%%%%%%%%%%%%%%%%%%%%%%%%%%%%%%%%%
%%%%%%%%%%%%%%%%%%%%%%%%%%%%%%%%%%%%%%

First of all, these solutions extend much further in $q$.  
At $q_{\rm max}$ the solutions are singular and possess zero area.
Second, the right outer boundary does not contain the static solution. 
Instead, two rotating $J=0$ solutions are encountered
at a value $q_1$ of the scaled charge that is still considerably larger than the static value.
Thus for the same global charges, there are two distinct (but symmetric) solutions
at $q_1$. 
Interestingly, however, the branches of extremal
black hole solutions also extend inside the domain of existence,
where they form an intriguing pattern of branches.
In particular, a whole sequence of rotating $J=0$ solutions arises.
%approaching the static solution.
Found at $q_2,q_3,\dots$ , all these solutions come in pairs with the same
global charges.

Investigating this sequence of extremal $J=0$ solutions for fixed charge $Q$
in more detail
we realize, that these solutions constitute a set of radially excited extremal solutions,
that can be labeled by the node number $n$ of the magnetic gauge potential $a_k(r)$,
as seen in Fig.~\ref{fig2},  or  equivalently, of the metric function $\omega(r)$.
% \cite{note1}. 
The first node always refers to spatial infinity.
We have constructed solutions with up to 30 nodes; thus it is likely that they form an infinite sequence.

\begin{figure}[t!]
\begin{center}
\mbox{\hspace*{-.8cm}
\includegraphics[height=.35\textheight, angle =270]{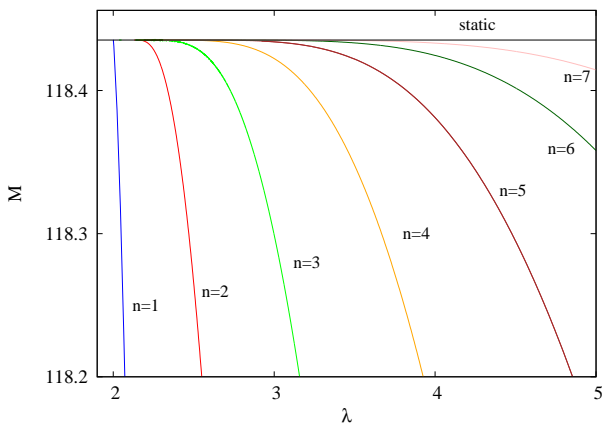}
}
\end{center}
\caption{\small
Mass $M$ versus $\lambda$ for rotating $J=0$ solutions
with several nodes.
}
\label{fig6}
\end{figure}

With increasing node number $n$,
the mass $M_n$ converges monotonically from below to the mass of the extremal static black hole, 
$M_{\rm RN}$  (see  Fig.~\ref{fig3}).
%This is seen in Fig.~\ref{fig2} (left),
%where the mass $M$ of the extremal solutions is shown versus the angular momentum $J$
%for fixed charge $Q$ at $\lambda=5$.
%The Smarr formula implies, that the horizon electrostatic potential of the sequence
%of extremal $J=0$ solutions converges likewise to the extremal static value. 
%At the same time, the horizon angular velocity converges to zero.
Surprisingly, however, the horizon area has the same value
for all solutions of the sequence and the same holds for the 
magnitude of the horizon angular momentum.
For comparison, we exhibit in Fig.~\ref{fig4}
the area versus the angular momentum as obtained
within the near horizon formalism for the same charge $Q$
% and the same CS coupling $\lambda=5$. 
(Figs.~\ref{fig2}-\ref{fig6} are for a given value 
%$Q=136.75$;
$Q=8 \sqrt{3} \pi^2$;
 for completeness, the charge $-Q$ is also shown).

The set of near horizon solutions exhibits only three $J=0$ solutions, the
extremal static one and two (symmetric) rotating solutions with lower area.
The latter have precisely the values found for all solutions of the sequence.
Thus these two near horizon solutions are not only associated with  a single global solution each,
but with whole sequences of global solutions.

Let us next address the full set of extremal solutions for fixed charge $Q$
inside the domain of existence.
As expected, sequences of radially excited  solutions \cite{non-extr} 
%labeled by $n$
exist also for  nonzero $J$. As seen in Fig.~\ref{fig3}
two (symmetric) $n=1$ branches of solutions extend from the $n=1$ $J=0$ solutions
up to a local minimum resp.~maximum of the angular momentum. 
There cusps $C_{1,2}$ are encountered formed with higher mass $n=2$ branches.
These higher mass branches then pass the $n=2$ $J=0$ solutions and end
at bifurcations $B_{1,2}$ with the opposite $n=1$ branches.

%%%%%%%%%%%%%%%%%%%%%%%%%%%%%%%%%%%%%%%%%%%%
%%%%%%%%%%%%%%%%%%%%%%%%%%%%%%%%%%%%%%%%%%%%

%%%%%%%%%%%%%%%%%%%%%%%%%%%%%%%%%%%%%%%%%%%%
%%%%%%%%%%%%%%%%%%%%%%%%%%%%%%%%%%%%%%%%%%%%

Considering this same set of solutions in Fig.~\ref{fig4}
we see, that the cusps $C_{1,2}$ encountered by the global
solutions do not correspond to the cusps $C_{\rm nh}$ of the solutions
obtained in the near horizon fomalism, since the branches of global solutions
end before the branches of near horizon solutions end.
Thus there are sets of near horizon solutions which do not possess
counterparts in the sets of global solutions \cite{Galtsov}.

The bifurcations $B_{1,2}$
of the $n=2$ branches with the opposite $n=1$ branches are clearly
visible in Fig.~\ref{fig5}, 
where we exhibit the horizon angular velocity $\Omega_{\rm H}$
of the extremal solutions versus the angular momentum $J$.
On the $n=2$ branches bifurcations $B_{2,3}$
with $n=3$ branches arise. Passing
the $n=3$ $J=0$ solutions, 
these branches end in cusps $C_{3,4}$, formed with branches
of the $n=4$ solutions. 
The $n=4$ branches pass the $n=4$ $J=0$ solutions
and end in bifurcation points $B_{3,4}$ on the $n=3$
branches. 
This pattern then repeats again and again for the higher node branches.

In this way a whole sequence of branches is generated.
%with bifurcation points $B_{l,l+1}$ at constant $|J|$.
Since the cusps $C_{n,n+1}$ occur at decreasing values of $|J|$,
the number of extremal global solutions for fixed $|J|$
decreases with increasing $|J|$, whenever a cusp is passed.
We conclude that a given near horizon solution
can correspond to 
{\it i) more than one global solution}
(possibly even an infinite set),
{\it ii)  precisely one global solution},
or {\it iii)  no global solution at all}.

Interestingly, the presence of the bifurcation points $B_{n,n+1}$
indicates, that at each of these points there are two distinct solutions
with the same global charges. 
They possess, however, different values of the area
and thus correspond to different near horizon solutions.

Let us finally address the $\lambda$ dependence of the observed
pattern of extremal solutions.
%Clearly, there exists a critical value $\lambda_{\rm cr}$,
%beyond which no rotating $J=0$ solutions exist.
%At this critical value, the mass of the rotating $J=0$ solution and
%the extremal static solution should coincide,
%in order to have a smooth emergence of the new boundary point
%of the domain of existence at $\lambda_{\rm cr}$.
%
This is shown in Fig.~\ref{fig6} 
where we exhibit the mass $M$
versus the CS coupling $\lambda$ of the rotating $J=0$ solutions
with one to seven nodes.
We note, that we do not find rotating $J=0$ solutions below
$\lambda=2$, and the mass $M_n$ of the solutions with $n$ nodes
approaches the mass $M_{\rm RN}$ of the static extremal solution,
as $\lambda$ is decreased towards $\lambda=\lambda_{SG}$.

%%%%%%%%%%%%%%%%%%%%%%%%%%%%%%%%%%%%%%%%%%%%%%%%%%%%%%%%%%%%%%%%%%%
%\section{Conclusions}
%%%%%%%%%%%%%%%%%%%%%%%%%%%%%%%%%%%%%%%%%%%%%%%%%%%%%%%%%%%%%%%%%%%
%%%%%%%%%%%%%%%%%%%%%%%%%%%%%%%%%%%%%%%%%%%%%%%%%%%%%%%%%%%%%%%%%%%%%%%%%%%%%%%
{\bf Further remarks.--}
%%%%%%%%%%%%%%%%%%%%%%%%%%%%%%%%%%%%%%%%%%%%%%%%%%%%%%%%%%%%%%%%%%%%%%%%%%%%%%%%
The results in this work show that 
the intuition  
based on known exact solutions  
cannot be safely applied in the general case. 
Working in EMCS theory, we have shown that beyond a critical value of the CS coupling $\lambda$,
for fixed charge $Q$,
a sequence of branches of extremal radially excited black holes arises.
To our knowledge, this is the first example of 
black holes with Abelian fields which form excited states reminiscent of 
radial excitations of atoms \cite{note2}. 
%To our knowledge, 
% This resembles also the case of black holes with
% non-Abelian gauge fields \cite{Volkov:1998cc}.

%Intriguingly, however, the associated near horizon solutions
%consist of only two branches of solutions.
%As one moves along the branches of global solutions,
%going to increasingly higher nodes,
%one moves up and down the two branches of near horizon solutions.
%
%Of particular interest is a set of rotating $J=0$ black holes.
%With increasing node number
%their mass converges to the mass of the extremal static black hole,
%while their horizon angular velocity tends to zero
%along with their magnetic moment.
%Their area, however, retains the same constant value,
%different from the static value.
%
Also, these black holes clearly illustrate 
that the relation between the global solutions and the near horizon solutions
may be rather intricate.
%Comparison of the global solutions and the near horizon solutions shows,
Since there are near horizon solutions that do not correspond to global solutions,
while other near horizon solutions correspond to one or more global solutions,
possibly even infinitely many.

As we decrease the CS coupling below the critical value, this intriguing
pattern of global solutions disappears. 
However, in the limit $\lambda \to 0$  when EM theory is obtained, 
we again encounter two branches of extremal solutions. 
The first branch is connected to the static RN black hole and
has horizon area $A_H=\pi 3^{3/4}J^2 Q^{-3/2}/\sqrt{2}+3^{1/4}\sqrt{2}Q^{3/2}/48\pi$.
The second branch originates at the Myers-Perry solution, 
and has the rather unusual property to possess an
entropy independent of the electric charge,
$A_H=J/2$.
Both branches are associated with near
horizon solutions, that are only partly realized as global configurations. 

Finally, we conjecture that extremal black holes with similar properties 
may also exist in other theories, in particular in a $D=4$ EM-dilaton theory.
\\

%%%%%%%%%%%%%%%%%%%%%%%%%%%%%%%%%%%%%%%%%%%%%%%%%%%%%%%%%%%%%%%%%%%
{\bf Acknowledgement}
%%%%%%%%%%%%%%%%%%%%%%%%%%%%%%%%%%%%%%%%%%%%%%%%%%%%%%%%%%%%%%%%%%%
We gratefully acknowledge support by the Spanish Ministerio de Ciencia e
Innovacion, research project FIS2011-28013, and by the DFG, 
in particular, the DFG Research Training Group 1620 ''Models of Gravity''. 
J.L.B is supported by the Spanish Universidad Complutense de Madrid.

%%%%%%%%%%%%%%%%%%%%%%%%%%%%%%%%%%%%%%%%%%%%%%%%%%%%%%%%%%%%%%%%%%%


\begin{thebibliography}{000}
\bibitem{numerics}
 We employ a collocation method for boundary-value ordinary
differential equations, equipped with an adaptive mesh selection procedure
\cite{COLSYS}.
Typical mesh sizes include $10^3-10^4$ points, 
the solutions having a relative accuracy of $10^{-8}$.
\end{thebibliography}
\end{document}